# Empirical measures of the largest amounts of magnetically-induced radius inflation in low-mass stars

D. J. Mullan & J. MacDonald

Department of Physics and Astronomy, University of Delaware, Newark DE 19716

**Abstract**

Access to precise empirical estimates of stellar radii in recent decades has revealed that the radii of certain low-mass stars are inflated relative to stellar structure predictions. The largest inflations are found in magnetically active stars. Although various attempts have been made to incorporate magnetic effects into stellar structure codes, a major source of uncertainty is associated with our lack of knowledge as to how the field strength varies inside the star. Here, we point out that a recent study of 44 eclipsing binaries in the Kepler field by Cruz et al. may enable us for the first time to set an upper limit $B_c$ on the field strengths inside the 88 stars in the sample. According to our magneto-convective model, the largest empirical inflations reported by Cruz et al. can be replicated if $B_c \approx 10$ kG inside stars with masses $\geq 0.65$ $M_\odot$. On the other hand, in lower mass stars, especially those with masses of $\leq 0.4$ $M_\odot$, our model predicts that the largest empirical inflations may require significantly stronger fields, i.e. $B_c \approx 100\text{-}300$ kG.

## 1. Introduction

An important paper by Leggett et al. (2000) appears to be the earliest to report on empirical radii of low-mass stars where the data have sufficient precision to allow for a meaningful discussion of the question: do empirical radii of low-mass stars agree with the predictions of standard stellar structure models? To be sure, the error bars on the empirical radii derived by Leggett et al. (2000) were relatively large (see their Figure 13) due in part to uncertainties in deriving the bolometric luminosity from ground-based spectra at wavelengths of 1-2.5 microns. In a first conclusion, Leggett et al. reported that many of the stars in their sample were found to have radii which are indeed consistent (within error bars) with the standard models: that is, for many stars, the standard models do not appear to be subject to serious errors. Significantly, *none* of the stars in their sample was found to have a radius which was empirically *smaller* than the standard models predicted. However, in 5 of their stars, the radii were found to be (in a statistically significant sense) *larger* than the standard model predictions. Leggett et al. (2000) speculated that these 5 stars might be previously unknown multiple star systems, although no evidence was provided in support of this speculation.

In the present paper, we will use the term "inflated" to describe low-mass stars with empirical radii which are (statistically speaking) larger than standard models predict.

In a subsequent discussion of the Leggett et al. results, Mullan & MacDonald (2001: hereafter MM01) pointed out that the 5 "inflated" stars reported by Leggett et al. had one well-defined empirical property in common: they are all listed in SIMBAD as either "flare star" or "variable star". These two categories are significant in the context of magnetic fields. Firstly, "flare stars" are known to owe their flaring activity to the presence of magnetic fields in the outer regions of the star (e.g. Mullan 1989): such fields can be subjected to tangling by turbulent motions of gas in the extensive convective envelope which is a key structural property of low-mass stars. Convective tangling of magnetic fields can at times generate outbursts of stored magnetic energy via the physical process of magnetic reconnection. Secondly, "variable stars" may owe their variability to the presence of cool darker regions on their surfaces ("star-spots"). These star-spots exist because strong (vertical) magnetic fields concentrated in localized patches in the outer regions of the star have the ability to inhibit convective flows in electrically conducting material, thereby leading to features on the stellar surface which are noticeably cooler (darker) than the immaculate (un-spotted) regions of the surface (e.g. Mullan 1974).

The fact that 5 inflated stars in the Leggett et al sample exhibited features which are almost certainly associated with magnetic effects of some kind led MM01 to suggest that radius inflation in low-mass stars *might* be due (somehow) to the presence of magnetic fields.

## 2. Incorporating magneto-convection into stellar convection models

We note that in a stellar structure code which is relevant for *non-magnetic* gas, the criterion for the onset of convective instability is the widely-used Schwarzschild criterion, namely, the vertical temperature gradient $\nabla$ must exceed the adiabatic gradient $\nabla_{ad}$. However, it has been known for decades that in the presence of magnetic fields, the Schwarzschild criterion no longer suffices as a physically-motivated criterion for the onset of convection, or more strictly, for the onset of *magneto-convection*. For example, Gough and Tayler (1966: hereafter GT66) developed the following criterion for the onset of magneto-convection:

$$\nabla > \nabla_{ad} + \delta$$

where $\delta = B_v^2/(B_v^2 + 4\pi\gamma P)$ is the "magnetic inhibition parameter". In the formula for $\delta$, $B_v$ is the vertical component of the local magnetic field, $P$ is the local gas pressure and $\gamma$ is the local ratio of specific heats. GT66 derived their criterion by using a physics-based energy principle which has been widely used in the study of laboratory plasma physics.

When GT66 developed their convective instability criterion, they were interested mainly in replicating the structure of relatively small cool regions on the surface of the Sun (sun-spots). They examined in detail a particular empirical model of a sunspot which had been obtained by Chitre (1963) and found that their criterion is satisfied throughout most of the empirical model. GT66 surmised that the inability to satisfy the criterion in *all* regions of Chitre's model arises because of omissions of certain physical processes from their "simple theory" (including boundary conditions, field curvature, finite transport processes). In what follows, we shall find that the presence of cool spots on stellar surfaces can only provide a partial solution to the problem of radius inflation in stars.

The simplest approach to incorporating magnetic effects into a stellar evolution code is to reduce the mixing length parameter associated with convection in the stellar envelope to values which are significantly smaller than usually assumed. This is the approach which was used by Gabriel (1969) and by Cox et al. (1981).

A theory of stellar magneto-convection which is more detailed and more complicated than that of GT66 was presented by Lydon & Sofia (1995: hereafter LS95). Included in the LS95 treatment are extensive changes to the thermodynamic properties of the stellar material, such as specific heats and adiabatic gradient. The magnetic "perturbations" which were introduced into the model were not global field structures: instead, they consisted of a magnetic field enhancement which is assumed to be localized within two narrow range of depths, the first (the deep perturbation) with field strength of $10^6$ G situated below the inner boundary of the convection zone, and the second (the shallow perturbation) with field strength of $10^3$ G near the surface. These perturbations were found to lead to slight fractional radius inflations $\Delta R/R$, with numerical values of $\Delta R/R = 5 \times 10^{-5}$ for the deep perturbation, and $\Delta R/R = 10^{-6}$ for the shallow perturbation.

### 3. Application of the GT66 criterion to low-mass stars

In a study of lithium depletion in cool dwarfs, Ventura et al (1998) applied the GT66 criterion to the evolution of pre-main sequence stars.

In order to study whether or not magnetic effects might contribute to stellar radius inflation in low-mass stars on the main sequence, MM01 also applied the GT66 criterion.

The major unknown in applying the GT66 criterion to the computation of a model for a magnetic star is the following: how does the vertical field strength $B_v$ vary as a function of radius inside the star? Finding an answer to this question is not a simple matter. Empirical information on the radial magnetic profile $B_v(r)$ might in some circumstances be extractable from asteroseismic data if eigenfrequencies of multiple modes could be reliably identified. This approach has already proven useful in studying internal fields (with strengths of 13-150 kG) in red giants where *p*-modes in the outer envelope couple with *g*-modes in the core (e.g. Li et al. 2023), although the interpretation of the data can be complicated (Jiang et al. 2022). However, there is currently no proof that pulsations of any kind (either *p*-modes or *g*-modes) have been reliably identified in low-mass M dwarfs (e.g. Emmanuel-Martinez et al. 2023). Therefore, we cannot yet rely on asteroseismic information to assist us in specifying the radial profile of magnetic fields in the low-mass stars which are of interest to us here. For such stars, we are forced to choose a profile in a more or less *ad hoc* manner.

Initially, one of the magnetic profiles chosen by MM01 was based on setting the magnetic inhibition parameter $\delta$ equal to a constant numerical value throughout the star. But this is unsatisfactory because it leads to enormously strong fields (10's of MG) in the core of a low-mass star: it is not physically obvious how such strong fields could be generated. Moreover,

Browning et al (2016) have shown that field strengths in excess of roughly 1 MG would not be stable in low-mass stars.

In papers which we published subsequent to MM01, among the various possibilities which might be considered to model the physics of magneto-convection, we opted for the following approach: we *assume* that $\delta$ remains constant only in the *outer regions* of the star. In these outer regions, as we move inwards deeper into the star, the continual increase in gas pressure means that in order to keep $\delta$ = constant, the strength of $B_v$ must continually increase above its surface value as we go deeper into the star. However, at a certain depth, i.e. at a certain radial location, $r = r_c$, where the field strength $B_v$ has increased to a certain value $B_v(c)$, we *assume* that no further increase in $B_v$ occurs in the deeper interior of the star. We refer to $B_v(c)$ as the "ceiling field": hence the subscript $c$ attached to the radius at which this field is reached. At radial locations which lie deeper inside the star than $r = r_c$, we assume that $B_v$ remains constant. In other words, for $r < r_c$, $B_v(r)$ is assumed to remain equal to $B_v(c)$. Accordingly, as the pressure continues to increase inwards towards the stellar core, the local numerical value of the magnetic inhibition parameter $\delta$ will steadily *decrease* in numerical value as we move deeper into the star beneath the location $r = r_c$. As a result, magnetic inhibition effects will be maximally effective in the outer regions of the star $r > r_c$ where the most noticeable empirical phenomena (star-spots, flares) are most prominent.

When the GT66 formula is used, MM01 (and subsequent papers) found that the radius of a magnetic star is indeed "inflated" relative to the radius of a non-magnetic star of the same mass. In the years since MM01, certain studies of detached eclipsing binaries (DEB), have been successful in determining empirical stellar radii and masses with increasingly high precision. As a result, fractional radius inflations as small as $\Delta R/R < 1\%$ can now be confidently identified in the best cases. Between 2007 and 2024, generalizations of the MM01 approach, that include non-ideal effects, has been applied to replicate radius inflations which were observed in a variety of stars, ranging from the Sun to brown dwarfs (Mullan et al. 2007, MacDonald & Mullan 2024b). The amplitudes of the empirical values of $\Delta R/R$ in the stars we modelled ranged from a few percent to at most roughly 20%. In our studies, we found that it is possible to replicate (within the quoted error bars) as many as 20 empirical measures of radius inflation in lower main sequence stars by choosing a value of $10^4$ G for the 'ceiling field' (MacDonald & Mullan 2017).

4. **Effects of cool spots on the radii of low-mass stars**

As mentioned in Section 1 above, the presence of (vertical) magnetic fields occupying certain patches of the stellar surface (thereby blocking the outward flow of convective energy in those patches) requires re-distribution of energy outflow in the star. This re-distribution process can also lead to radius inflation (e.g. MacDonald & Mullan 2012). This raises the question: of the two distinct physical mechanisms which can produce radius inflation (magneto-convection, spots), is it possible to identify an empirical criterion which can help us to decide whether one or other mechanism might be preferable in a particular star?

In this regard, the system EPIC 219511354 is a binary consisting of two K stars in which the masses and radii of both components have been obtained with high precision (1-2%). We have shown that the magneto-convective model can fit the radii of both components satisfactorily within the (small) 1 $\sigma$ error bars (Mullan & MacDonald 2022). On the other hand, Torres et al (2022) have also shown that the empirical radii of both components can be fitted satisfactorily with a star-spot model. However, in order to fit the secondary star in this system, the star-spots are required to occupy a fractional surface area of up to 88-94% (Torres et al. 2022).

These results raise the physics issue as to how a star would generate a magnetic field which impedes convection effectively over as much as 94% of the surface area. Such a star would have to emit the majority of its radiant energy from only 6% of its surface.

## 5. Predicting upper limits on radius inflation in the magneto-convection model

The magneto-convection model allows us to examine quantitatively the following question: does there exist any theoretical upper limit on the amount of radius inflation? MacDonald and Mullan (2024a) have found that in certain cases, the answer to this question is Yes. Thus, for a ceiling field of 10 kG, the maximum theoretical value of the fractional radius inflation $\Delta R/R$ is predicted to be close to 100% for stars of mass 0.7 $M_\odot$. If the ceiling field is 100 kG, maximum inflations are predicted to be about 130% (also in stars of mass 0.7 $M_\odot$). If the ceiling field is assumed to be as large as $10^6$ G (the largest value which is compatible with stellar stability: see Browning et al. 2016), inflations as large as 350% are predicted to occur in stars of mass 0.9 $M_\odot$.

These results naturally lead to the question: are there any empirical data which might allow our predicted upper limits on radius inflation to be tested? We turn to a consideration of data which may be of assistance in this regard.

## 6. Comparison between the largest empirical values of radius inflation and the theoretical limits predicted by magneto-convection

Cruz et al. (2022) have reported on a process which allowed them to extract stellar temperatures, radii, and masses by using photometric data for 88 components of detached eclipsing binaries in the Kepler field. (In the notation of Cruz et al, the 88 components belong to their "sub-sample B".) Using evolutionary models computed by Bressan et al (2012), Cruz et al selected dwarf star models with $T_{eff}$ in the range 2303-5991 K and they created a grid of synthetic model binaries containing almost 200,000 entries, for each of which they computed 10 different colors based on adjacent filters used by PanSTARRS (*g,r,z,i,y*) and by 2MASS (*J,H,K*). The observed photometric data in all 10 colors were then interpolated into the grid to determine for each star its $T_{eff}$. At this stage, Cruz et al used a table of $T_{eff}$ and masses compiled by Pecaut & Mamajek (2013) to extract a "photometric mass" for each star. With $T_{eff}$ values in hand, the light curve of each binary was analyzed photometrically to extract the sum of the radii of the two stars and the ratio of the two radii: combining these quantities, Cruz et al obtain "photometric radii" of primary and secondary components.

We compare radius predictions *from our magneto-convection models* with data from Cruz et al. (2002) in the left-hand panel in Figure 1. The black squares show the largest values of fractional radius inflation reported by Cruz et al at various stellar masses. The broken and solid black lines show the predicted radii for models of two different ages without magnetic field: for example, note that a non-magnetic star with mass 0.7 $M_\odot$ is predicted to have a radius of 0.6-0.7 $R_\odot$. The colored lines show the predicted isochrones of our magneto-convective models for different sets of values of $\delta$, $B_v(c)$ and age. We have chosen $\delta = 0.4$ as a representative value of the magnetic inhibition parameter (see MacDonald & Mullan 2024 for the effects of choosing a different value for $\delta$). Note that for a star of mass 0.7 $M_\odot$, the radius predicted by a magneto-convection model with $B_c = 10$ kG is about 1.2 $R_\odot$, i.e. about twice as large as the non-magnetic star with the same mass. Thus, the fractional radius inflation $\Delta R/R$ in this case is close to 100%. This value is not far from the maximum empirical inflation reported by Cruz et al. (2022).

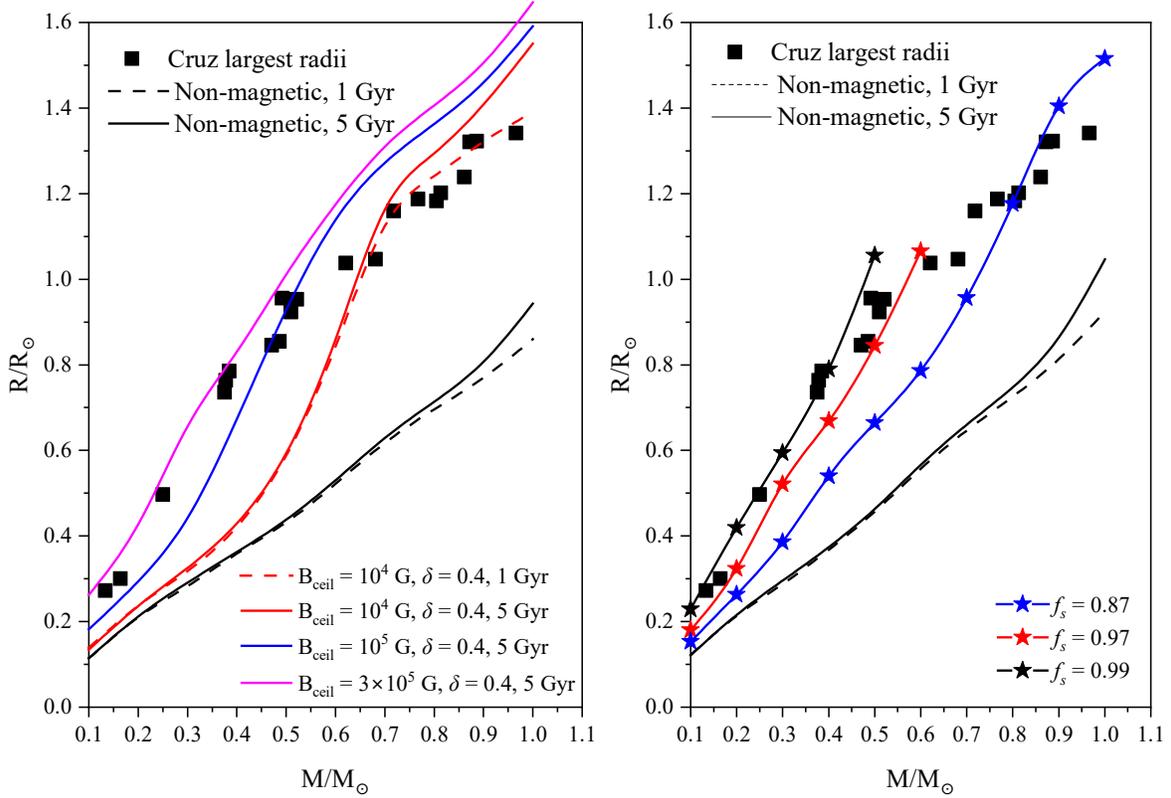

We see that, according to our magneto-convection models of stars with masses greater than about 0.65 $M_\odot$, the largest observed inflations can be achieved with $B_v(c) = 10$ kG. Such a value for the "ceiling" field has already emerged in our earlier work on replicating empirical radii in some 20 stars on the lower main sequence (see Section 3 above).

However, when we examine stars with smaller masses, a different value of the ceiling field emerges: $B_v(c)$ must be of order 100 kG to replicate the largest observed inflations in stars of mass between ~0.45 and ~0.65 $M_\odot$. And for stars of mass less than ~0.45 $M_\odot$, an even stronger ceiling field, $B_v(c)$ = 300 kG, is required to replicate the largest observed inflations.

*Inclusion of a "photometric" caveat:* in their photometric analysis, Cruz et al made use of evolutionary tables (Bressan et al. 2012) which were computed for *"standard" (non-magnetic) stellar models*. However, we are dealing here with *magnetic* stars, and for these, the surface temperatures can be altered from the Bressan et al. values by magnetic effects. (e.g. MM01) As a result, the 10-color system which is suitable for deriving photometric properties of non-magnetic stars needs to be corrected if the colors are to be applied to magnetic stars. Specifically, Mullan & MacDonald (2022) pointed out that the "photometric masses" derived by Cruz et al (2022) are actually *lower limits* on the actual masses. But the corrections in mass are highly non-linear: stars with masses of 0.65-0.71 $M_\odot$ need the smallest corrections, and these are the stars which are predicted to undergo the maximum inflation (MacDonald & Mullan 2024). On the other hand, stars with masses of 0.13-0.39 $M_\odot$ require larger corrections in "photometric mass". Thus, some of the estimates of radius inflation reported by Cruz et al. need to be reduced, especially in the range of masses which lie close to the transition to complete convection on the main sequence.

To the extent that the photometric mass corrections can be considered to be relatively small, it is tempting to consider that the onset of complete convection in main-sequence stars (i.e., stars have masses less than roughly 0.35 $M_\odot$) may play a role in explaining why stronger ceiling fields are required to replicate the largest empirical inflations in stars with the lowest masses in Figure 1 (left panel). Completely convective stars rely on distributed dynamos operating throughout the star to generate their fields, while stars with radiative cores may also rely on a shell dynamo operating at the interface between radiative core and the outer convective envelope. If our conclusions regarding the strength of the ceiling fields mentioned above can be substantiated (i.e. weaker ceiling fields in more massive dwarfs, stronger ceiling fields in less massive dwarfs), it could suggest that the distributed dynamo operating throughout the entire volume of the star is more effective at generating fields than the interface dynamo.

### 7. Are there theoretical limits on radius inflation due to star-spots?

As mentioned in Section 4 above, star-spots and magneto-convection have both been found to be possible mechanisms to replicate radius inflation. Now that we have shown that upper limits exist on radius inflation in the case of *magneto-convection*, it is natural to ask: are there also upper limits on the radius inflation predicted by *star-spot* models?

Using the model for the effects of completely dark spots (i.e. with spot temperatures of zero) described by MacDonald & Mullan (2012), and comparing with the data of Cruz et al. (2022), we get the results shown in the right-hand panel of Figure 1. We find the radius inflations for stars with masses greater than about 0.8 $M_\odot$ are consistent with dark spot coverage factors of 87% or less. For lower mass stars, even larger dark spot coverage factors are needed to explain the maximum observed inflations: a coverage of 99% is needed for masses less than 0.4 $M_\odot$.

We stress that these coverage factors are for *completely dark* spots. If we relax this condition, we might allow the spots to have finite temperatures of 80% of the unspotted surface temperature (such as was assumed by Torres et al. [2022] in their study of the EPIC 219511354 secondary). But in such a case, we find that fits to the empirical inflations reported by Cruz et al. (2022) are impossible to achieve.

As mentioned in Section 4, it is difficult to envision from a physical perspective how a star could have spots which occupy as much as 99% of the surface area. In view of this, we consider it preferable to analyze radius inflations in low-mass stars in terms of a magneto-convective model instead of a spot model.

## 8. Conclusion

Increasing numbers of low-mass stars have been identified as having radii which are larger than standard stellar structure calculations predict. The occurrence of "radius inflation" is especially prevalent among stars which exhibit symptoms of magnetic activity. In view of this, we have been applying a model of magneto-convection in order to replicate the empirical inflations.

In 2024, we used our model to predict theoretically the maximum amounts of magnetically-induced inflation in stars of various masses in the presence of certain upper limits on the magnetic field strength inside the star (see MacDonald & Mullan 2024). The upper limits we selected for the internal field strength were 10 kG, 100 kG, and 1 MG: the latter value is the strongest field which can exist stably inside a low-mass star (Browning et al. 2016). In the case of 10 kG fields, we found that the maximum value of the relative radius inflation would be $\Delta R/R \approx 100\%$ in a star of mass 0.7 $M_\odot$. In the presence of 100 kG fields, $\Delta R/R$ was predicted to reach about 130%. And in the limit of 1 MG, relative inflations of as much as 350% were predicted.

In 2022, Cruz et al. (2022) published empirical results of relative radius inflations $\Delta R/R$ for some 88 stars in Kepler data: in some of their stars, they concluded that the $\Delta R/R$ estimates could be as large as $\approx 100\%$. (See Section 6 above for a *caveat* regarding this conclusion.) In the present paper, we report on comparisons between these empirical results and our theoretical predictions. We find that for stars of mass $\geq 0.65$ $M_\odot$, maximum internal fields of 10 kG can replicate the maximum inflations reported by Cruz et al. (2022). This is consistent with earlier work of ours which suggested that a maximum field strength of 10 kG inside a star can replicate the inflations which have been reported in some 20 stars on the lower main sequence (MacDonald & Mullan 2017). However, in order to replicate the maximum inflations reported by Cruz et al. for stars of mass $\leq 0.4$ $M_\odot$, we find that the maximum field strength inside the star must be as large as 300 kG.

If these results can be confirmed, they will suggest that, other things being equal, a dynamo in a completely convective star may generate significantly stronger fields than a dynamo in a star consisting of a radiative core plus a convective envelope.

**Acknowledgements**

J.M. acknowledges partial support from the Delaware Space Grant Consortium through NASA National Space Grant NNX15AI19H.